\documentclass[twocolumn, switch]{article} 

\usepackage{preprint}
\usepackage{amsmath, amsthm, amssymb, amsfonts}
\usepackage{siunitx}
\usepackage[numbers,square]{natbib}
\usepackage[utf8]{inputenc}	
\usepackage[T1]{fontenc}	
\usepackage{xcolor}		
\usepackage[colorlinks = true,
            linkcolor = purple,
            urlcolor  = blue,
            citecolor = cyan,
            anchorcolor = black]{hyperref}	
\usepackage{booktabs} 		
\usepackage{nicefrac}		
\usepackage{microtype}		
\usepackage{lineno}		
\usepackage{float}			

\usepackage{lipsum}		

\usepackage{newfloat}
\DeclareFloatingEnvironment[name={Supplementary Figure}]{suppfigure}
\usepackage{sidecap}
\sidecaptionvpos{figure}{c}

\usepackage{titlesec}
\titlespacing\section{0pt}{10pt plus 3pt minus 3pt}{1pt plus 1pt minus 1pt}
\titlespacing\subsection{0pt}{10pt plus 3pt minus 3pt}{1pt plus 1pt minus 1pt}
\titlespacing\subsubsection{0pt}{8pt plus 3pt minus 3pt}{1pt plus 1pt minus 1pt}

\date{}

\title{\bf{Generating 500\,mW for laser cooling of strontium atoms by injection locking a high power laser diode}}

\usepackage{xwatermark}

\usepackage{authblk}

\author[1,*]{Vladimir Schkolnik}
\author[1]{Jason R. Williams }
\author[1]{Nan Yu}
\affil[1]{\normalsize  \textit{Jet Propulsion Laboratory California Institute of Technology, Pasadena, USA}}
\affil[*]{\normalsize  \textit {Now affiliated with: Humboldt-Universit\"at zu Berlin, Newtonstr. 15, 12489 Berlin, Germany, vladimir.schkolnik@physik.hu-berlin.de}}

\begin{document}

\twocolumn[ 
  \begin{@twocolumnfalse} 
  
\maketitle
\vspace{-0.15cm}
\begin{abstract}
We report on the generation of \SI{500}{mW} of spectrally pure laser light at the \SI{460.86}{nm} transition used for laser cooling of strontium atoms. To this end we inject a high power single mode laser diode with light from a stabilized  extended cavity diode laser. To optimize and monitor the injection status and the spectral purity of the slave diode we developed a novel technique that uses a single passive optical element. A narrow band interference filter generates a suitable monitoring signal without any additional electronics for post processing of the data. Our method greatly simplifies the daily operation of injection locked laser diodes and can be easily adapted to other wavelengths of interest.

\end{abstract}
\vspace{0.5cm}

  \end{@twocolumnfalse} 
] 



\section{Introduction}
\label{sec:intro}

Laser cooled strontium atoms are used as an atomic reference for high precision and accurate optical lattice clocks \cite{Bothwell_2019,Norcia2019,Madjarov2019}, quantum simulation and computing \cite{Norcia2018,Madjarov} and were recently proposed to detect gravitational waves \cite{PhysRevA.96.042118,PhysRevD.94.124043} and search for dark matter \cite{Derevianko_2016}. Mobile strontium lattice clocks would have far-reaching applications like  GPS, time scales and relativistic geodesy \cite{Grotti2018,McGrew2018}. For laser cooling of strontium at the \SI{461}{nm} transition laser diodes became only recently available. Therefore most setups so far traditionally employed relatively complex systems based on second harmonic generation (SHG) delivering output powers up to \SI{600}{mW} \cite{Barbiero2020}. 

As an attractive alternative to SHG systems, direct diode lasers at \SI{461}{nm} with output powers in the range of \SI{100}{mW} appeared as a commercial product several years ago (Nichia, NDB4216). This enabled using only a fraction of the power from a single frequency master laser that can be amplified by a slave high power laser diode  through the injection locking technique \cite{Hadley1986}. Several slave diodes, in some cases 10 or even more, can be used if needed to achieve the required output power \cite{Shimada2013}. Such amplified diode laser systems were already developed for laser cooling of Ytterbium at \SI{399}{nm} \cite{Hosoya2015} while some setups needed to operate with a much higher number of slave diodes \cite{Yu2006}.  Ideally one would use a single diode laser in combination with an amplifier chip in the so called MOPA configuration. Such chips are in development but they are not commercially available yet at the moment \cite{Stanczyk:18,10.1117/12.2286044}.  

Here we report the generation of \SI{500}{mW} single frequency laser light by injection locking a high power single transverse mode laser diode with a master laser at \SI{460.86}{nm}. The obtained output power is comparable to commercial available SHG systems \cite{Barbiero2020} in a simpler setup with lower total electric power consumption for a fraction of the cost, while being also more compact. For the analysis and optimization of the injection locking we use a single optical element, a narrow band interference filter (IF). The power transmitted through the filter generates a signal with a high signal to noise that indicates the quality of the injection locking state. The signal can also be potentially used for active stabilization of the injection lock \cite{Saxberg2016}. We analyze the injection lock behaviour for different slave currents and injection powers from the master laser. The demonstrated high power injection locked laser approach is universal and can be used in a variety of experiments utilizing laser cooled strontium atoms and other setups where injection locking is also utilized.

\section{Experimental setup}
\label{sec:setup}

\begin{figure}[htb]
\centering
\includegraphics[width=0.99\linewidth]{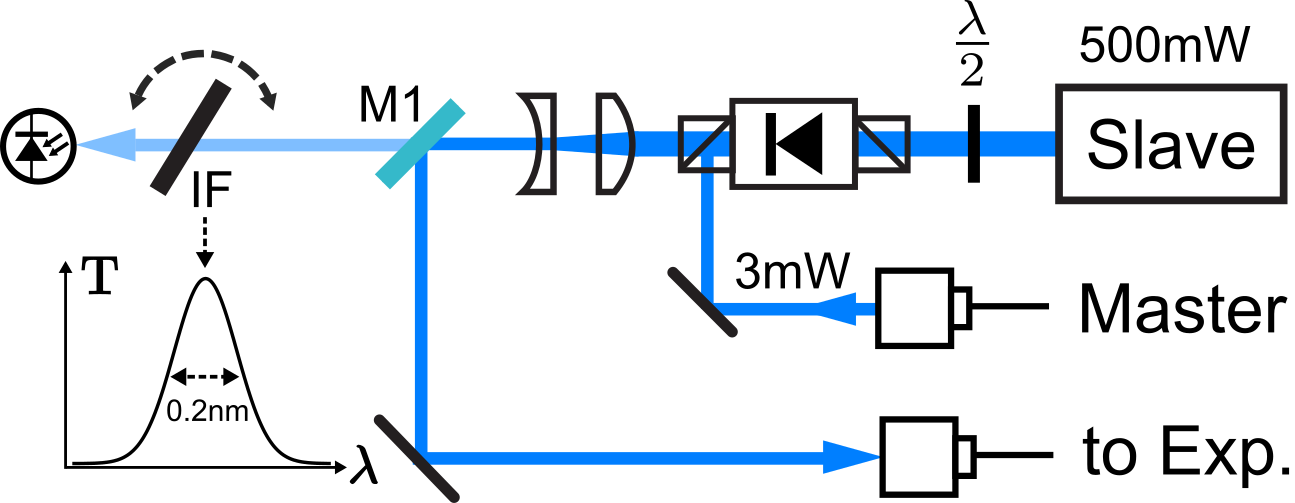}
\caption{A schematic of our setup. Light from the master laser is coupled into a fibre and injects a high power slave laser diode with an output power of \SI{500}{mW}. A small fraction of the slave  beam is transmitted through a backside polished mirror (M1) and is used for monitoring of the injection lock quality using the transmitted power through the interference filter (IF). See main text for more details.  }
\label{fig:design_ECDL}
\end{figure}

The experimental setup for the injection locking and characterization is shown in Figure \ref{fig:design_ECDL}. The slave laser is a single mode high power GaN laser diode (Nichia NDB4916E) that recently became available. The sample diode has a free running center wavelength around \SI{460.5}{nm} and an output power of \SI{500}{mW} at its nominal operation current of \SI{338}{mA}. The diode is housed in a collimation tube and the output is collimated with an aspheric lens ($NA = 0.6$). Light from the diode passes through an optical isolator with Polarizing Beamsplitting Cubes (PBS) on both sides and  $\lambda/2$ wave plate in front of the isolator maximizes the transmission to \SI{86}{\percent}. An optical fibre delivers up to \SI{3}{mW} from the master laser  to the setup. This light is injected into the slave diode using the rejection port of the optical isolator as shown in Figure \ref{fig:design_ECDL}. 

To maximize the spatial mode overlap between the master and the slave we couple first light from the slave in an optical fibre. Afterwards we turn the slave current off and couple the master light reflected from the slave front facet into the same optical fibre. The slave laser mode is elliptical with a 1:3 aspect ratio. A cylindrical lens pair after the isolator (see Figure \ref{fig:design_ECDL}) is used to circularize the high power slave laser output beam for better efficiency for fiber coupling.

The output from the slave is coupled into the main fibre that goes to the switching and distribution board with a coupling efficiency of \SI{65}{\percent}. Two mirrors are used to maximize the fibre coupling. One of the mirrors has a polished backside (labeled M1 in Figure \ref{fig:design_ECDL}). This mirror transmits approximately \SI{1}{\percent}  of the slave light. A narrow interference filter (IF) with a bandwidth (FWHM) of \SI{0.2}{nm} at \SI{463}{nm} designed at the incident angle of \ang{0} is placed after this mirror. The filter is mounted into a precision rotation stage. The transmitted slave power is monitored by an amplified photo diode. The filter transmission is angle dependent \cite{Baillard2006} and the filter angle is rotated to maximize the transmission of the locked master when the slave is turned off. The optical spectrum of the slave is analyzed with a fibre coupled  grating-based laser wavemeter with a spectral resolution of \SI{0.02}{nm} was placed either before or after the IF for the following measurements.

\subsection{High power diode spectrum analysis}
\label{subsec:XXX}

\begin{figure}[htb]
\centering
\includegraphics[width=0.99\linewidth]{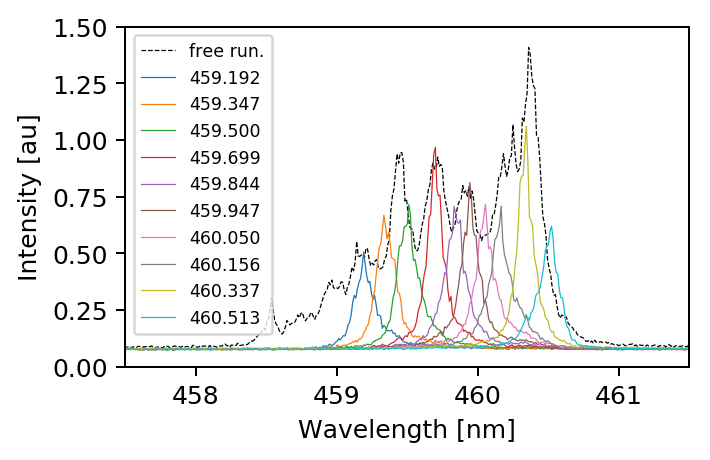}
\caption{Transmission spectra of the free running laser diode at an injection current of \SI{338}{mA} (dotted grey) and for different interference filter angles and thus transmission wavelengths. (solid colors). One clearly sees that the  filter is able to transmit only a small portion of the spectrum, and strongly suppress the rest.}
\label{fig:filter}
\end{figure}

 First we explore the filtering properties of our IF filter. The optical spectrum of the free running slave laser diode at an injection current of \SI{338}{mA} is recorded with the wavemeter. This spectrum is shown in  in Figure \ref{fig:filter} (grey dashed) and was taken without the IF, while all other spectra in Figure \ref{fig:filter} were taken with the IF inserted into the beam path. Since the bandwidth of our IF filter is much narrower than the spectrum of the free running slave laser diode, we change the filter angle in small increments to tune the center of the filter transmission. Figure \ref{fig:filter} shows the filtering performance of the interference filter for the the entire tuning range of filter angles spanning approximately \SI{2}{nm}. It is immediately apparent that the \SI{0.2}{nm} wide IF filter can reliably transmit a narrow band from the full slave spectrum.
 
 Several sources in the literature report that GaN diodes have internal modes separated $\approx$  by \SI{0.05}{nm} \cite{Martin2016}. Some of the filtered spectra in Figure \ref{fig:filter} reveal multiple closely spaced modes, but they cannot be reliably resolved with the spectrometer resolution.

\subsection{Narrow band interference filter for monitoring the injection lock of a laser diode}
\label{subsec:XXXXX}

\begin{figure}[htb]
\centering
\includegraphics[width=0.95\linewidth]{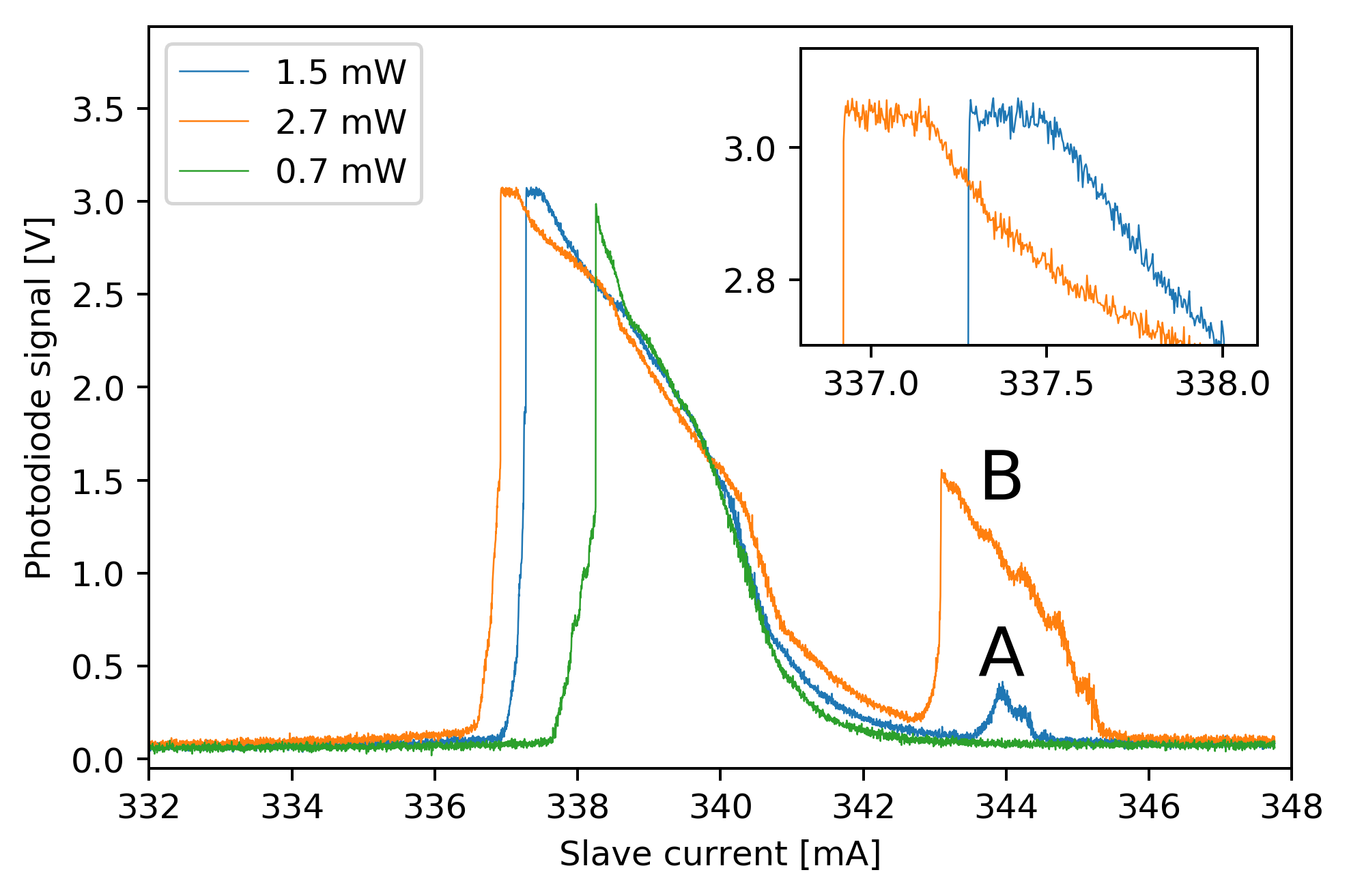}
\caption{The photo diode signal after the IF obtained by scanning the slave diode current by \SI{20}{mA} at a rate of \SI{100}{Hz} for three different injection power levels from the master. Partially injected regions are marked with the  A and B. The inset zooms on the plateau region, the current range where the slave is fully injection locked. The master laser was locked to an atomic transition during this measurement. (See main text for more details)}
\label{fig:osci_low}
\end{figure}

While optimizing and monitoring the quality of the injection lock can be done with the wavemeter alone, or like already demonstrated with a scanning Fabry–Pérot cavity \cite{Hirano2005}.  These devices are relatively complex, costly and require additional driving electronics. Especially in the case of Fabry–Pérot  cavities the transmission signal can show one single peak only, but not necessarily all the power of the slave is in the desired frequency mode \cite{Saxberg2016}.  Utilizing an IF that combines the filtering qualities of a filter cavity, but with a higher bandwidth, no scanning and/or fitting of signal is necessary and the power transmitted through the IF generates a signal directly proportional to the optical power in the desired mode of the slave diode. This will be presented and analyzed in this subsection.

The usefulness of the IF filter as an injection lock monitor relies on a precise alignment of the filter angle, so that the transmission is centered on the optical wavelength of the locked master. To achieve this, we first repeat the procedure described in Section \ref{sec:setup} to ensure that the beam overlap is optimized. Then we use the master light transmitted through the back polished mirror and optimize the IF filter angle for maximum transmission and therefore maximum photo diode signal. We want to emphasize, that the beam pick off for the IF filter monitor can be installed anywhere in the beam path of the slave. The IF and its photo diode detector can be made very compact and therefore included non-invasively almost anywhere in the beam path. Either after a fibre, in a unused beam splitter port, or inside one of the fibre collimators  that deliver light to the experiment.

 To explore the injection dynamics of the slave diode we scan the slave current \SI{20}{mA} around its operation point at a rate of \SI{100}{Hz} and observe the photo diode signal behind the IF filter with an oscilloscope. The obtained photo diode signals for different master injection powers are shown in Figure \ref{fig:osci_low}. The signal for an injection power of \SI{0.7}{mW} is shown in  Figure \ref{fig:osci_low} in green. The photo diode value and therefore the power of the master frequency component in the slave spectrum increases when the slave current is scanned from a higher current to a lower one. There is no region where the photo diode power is constant over some slave current range for this low injection power. Such a power plateau would have indicated that the slave is fully injection locked (meaning that the slave laser operates at the same frequency as the master) \cite{Saxberg2016}.

 For \SI{1.5}{mW} (blue curve in Figure \ref{fig:osci_low}) such a plateau can be seen at a slave current around \SI{337.5}{mA}. In addition a small second region where the slave diode is partially injection locked is visible (marked with the letter A). Such partially injection locked regions were already reported \cite{Hosoya2015}. For \SI{2.7}{mW} a \SI{0.3}{mA} wide plateau around \SI{337.0}{mA} can be seen seen where the slave diode is fully injection locked (orange curve in Figure \ref{fig:osci_low}) along with a second partially injected region (marked with the letter B). An inset in Figure \ref{fig:osci_low} enlarges the plateau region. All these measurements were carried out and verified simultaneously with the constant monitoring of the wavemeter.
 
 The monotone increase of the photo diode signal allows for a side of fringe lock very close to the plateau using the slave current as an actuator and stabilizing the slave injection lock fraction slightly below \SI{100}{\percent}. Alternatively one can recenter the slave current to the centre of the plateau at regularly time intervals \cite{Saxberg2016,Yu2006} since the injection lock is stable over long period of time and only occasional centering is needed. We want to indicate a shift of the injection lock region to smaller slave currents as a function of injection locked power as shown in Figure \ref{fig:osci_low}. We believe this is mostly due to the frequency pulling effect and resulting hysteresis.

\subsection{Injection locking quality and stability}

\begin{figure}[htb]
\centering
\includegraphics[width=0.99\linewidth]{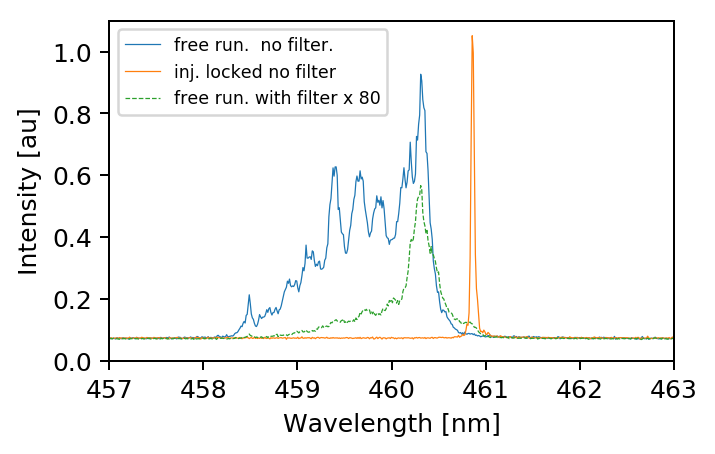}
\caption{Measurements of spectra with and without filter at \SI{338}{mA} injection current. The blue curve shows the free running laser diode without the IF in the beam path. The green curve is measured under the same conditions but with the filter inserted (Since the optical attenuation of the IF is very high outside its pass band the exposure time of the wavemeter was increased by a factor of 80 to obtain a visible signal). The orange curve shows the fully injected slave diode (recorded without a filter).}
\label{fig:free_running}
\end{figure}

The injection locking performance of the slave laser was also investigated with a wavemeter.  Figure \ref{fig:free_running} shows three different spectra. The free running diode without a filter is shown in blue. One can clearly see that the free running wavelength is centered around \SI{459.5}{nm} with no significant components at \SI{460.86}{nm} (the strontium cooling transition).  The same spectrum, but with the IF inserted into the beam path is shown in Figure \ref{fig:free_running} in green. Since the filter strongly suppresses the spectrum of the free running slave diode the exposure time of the wavemeter was increased by a factor of 80 to obtain a visible signal. The unwanted slave modes are strongly filtered in this case. 

The spectrum of the fully injection locked slave diode measured without any filtering from the IF is shown in Figure \ref{fig:free_running} in orange. All the integrated power of the free running slave diode is concentrated in this one optical mode. In this case only one optical mode,  the mode of the master laser is present and no other modes could be resolved. In addition we observe that the height of the peak detected by the wavemeter is at its maximum, showing that the optical power is not distributed over additional modes \cite{Saxberg2016}.  Other slave modes are suppressed in this case and a maximum IF transmission corresponds to a maximal slave injection quality. Even though the slave laser free-running frequency is more than \SI{1}{nm} away, the injection can pull the frequency and achieve stable locking. On the other hand, if the free running laser wavelength coincides with the master, the injection locking can be more stable and efficient in general. This could be changed in the future by a selected diode with a wavelength that matches closer the one of the maser. Alternatively one could heat the slave diode (\SI{1}{nm} increase in the center frequency requires a temperature increase by \SI{20}{K}) to the expense of a potential shortened lifetime. 

It should be pointed out that although the free running diode spectrum has no visible frequency components at the master seed frequency, we are still able to achieve full slave injection.  Optimizing the heat transfer in the diode mount and acoustic shielding of the free beam path between master and slave should lead to a better long term stability of the lock \cite{Saxberg2016}. Nevertheless after turning off the current scan and careful adjustment of the slave current, the slave stays fully injection locked over multiple hours without any intervention. This is worth mentioning due to the fact that our setup is located in a fairly busy lab and is not shielded against any disturbances in the environment.

\section{Conclusion}
\label{sec:conclusion}

 We generated \SI{500}{mW} output power of a spectrally pure single frequency in the blue region of the spectrum from a single laser diode  by injecting the diode with light from an ECDL master laser locked to an atomic transition in Strontium-88. The required power needed to achieve a full injection in our case is with less than \SI{3.0}{mW}.  Utilising the wavelength selective transmittance of a narrow band IF filter allows us to generate a signal suitable for optimizing and monitoring the injection lock quality, as well as provides a possible method for servo and maintain the injection locking for long period of continuous operation.
 
  A slave laser diode mount designed for better thermal and mechanical stability should improve the long term stability of the lock and a slave laser diode with a free running wavelength closer to the one of the master might lead to improved performance as well.
  
  These output power levels are en par with more complex, larger and expensive systems based on SHG generation. The low output power required from the master, allow to inject multiple slave laser diodes with a single master. For experiments where higher laser power is needed, this is a simple and effective solution. 
  
  Using the injecting master and slave laser configuration presented here, a complete strontium optical lattice clock  could be operated in the future \cite{Schkolnik2020}. This would  greatly reduce the complexity, size, power consumption and robustness of such a clock.

\section{Funding}
This work was carried out at the Jet Propulsion Laboratory, California Institute of Technology, under a contract with the National Aeronautics and Space Administration $\copyright$ 2020. California Institute of Technology. Government sponsorship acknowledged.

\section{Acknowledgment}

We thank Sean Mulholland, Matteo Sbroscia and Andrey Matsko for fruitful discussions, Klaus D{\"{o}}ringshoff for fruitful discussion and proofreading the manuscript.



\normalsize
\bibliography{ms}

\begin{thebibliography}{22}
\providecommand{\natexlab}[1]{#1}
\providecommand{\url}[1]{\texttt{#1}}
\expandafter\ifx\csname urlstyle\endcsname\relax
  \providecommand{\doi}[1]{doi: #1}\else
  \providecommand{\doi}{doi: \begingroup \urlstyle{rm}\Url}\fi

\bibitem[Bothwell et~al.(2019)Bothwell, Kedar, Oelker, Robinson, Bromley, Tew,
  Ye, and Kennedy]{Bothwell_2019}
Tobias Bothwell, Dhruv Kedar, Eric Oelker, John~M Robinson, Sarah~L Bromley,
  Weston~L Tew, Jun Ye, and Colin~J Kennedy.
\newblock {JILA} {SrI} optical lattice clock with uncertainty of $2.0 \cdot
  10^{-18}$.
\newblock \emph{Metrologia}, 56\penalty0 (6):\penalty0 065004, Oct 2019.
\newblock \doi{10.1088/1681-7575/ab4089}.

\bibitem[Norcia et~al.(2019)Norcia, Young, Eckner, Oelker, Ye, and
  Kaufman]{Norcia2019}
Matthew~A. Norcia, Aaron~W. Young, William~J. Eckner, Eric Oelker, Jun Ye, and
  Adam~M. Kaufman.
\newblock {Seconds-scale coherence on an optical clock transition in a tweezer
  array}.
\newblock \emph{Science}, 366\penalty0 (6461):\penalty0 93--97, oct 2019.
\newblock ISSN 10959203.
\newblock \doi{10.1126/science.aay0644}.

\bibitem[Madjarov et~al.(2019)Madjarov, Cooper, Shaw, Covey, Schkolnik, Yoon,
  Williams, and Endres]{Madjarov2019}
Ivaylo~S. Madjarov, Alexandre Cooper, Adam~L. Shaw, Jacob~P. Covey, Vladimir
  Schkolnik, Tai~Hyun Yoon, Jason~R. Williams, and Manuel Endres.
\newblock {An Atomic-Array Optical Clock with Single-Atom Readout}.
\newblock \emph{Physical Review X}, 91\penalty0 (4):\penalty0 041052, dec 2019.
\newblock ISSN 21603308.
\newblock \doi{10.1103/PhysRevX.9.041052}.

\bibitem[Norcia et~al.(2018)Norcia, Young, and Kaufman]{Norcia2018}
M.~A. Norcia, A.~W. Young, and A.~M. Kaufman.
\newblock {Microscopic Control and Detection of Ultracold Strontium in
  Optical-Tweezer Arrays}.
\newblock \emph{Physical Review X}, 8\penalty0 (4):\penalty0 041054, dec 2018.
\newblock ISSN 21603308.
\newblock \doi{10.1103/PhysRevX.8.041054}.

\bibitem[{Madjarov} et~al.(2020){Madjarov}, {Covey}, {Shaw}, {Choi}, {Kale},
  {Cooper}, {Pichler}, {Schkolnik}, {Williams}, and {Endres}]{Madjarov}
Ivaylo~S. {Madjarov}, Jacob~P. {Covey}, Adam~L. {Shaw}, Joonhee {Choi}, Anant
  {Kale}, Alexandre {Cooper}, Hannes {Pichler}, Vladimir {Schkolnik}, Jason~R.
  {Williams}, and Manuel {Endres}.
\newblock {High-Fidelity Control, Detection, and Entanglement of Alkaline-Earth
  Rydberg Atoms}.
\newblock \emph{arXiv e-prints}, art. arXiv:2001.04455, January 2020.

\bibitem[Norcia et~al.(2017)Norcia, Cline, and Thompson]{PhysRevA.96.042118}
Matthew~A. Norcia, Julia R.~K. Cline, and James~K. Thompson.
\newblock Role of atoms in atomic gravitational-wave detectors.
\newblock \emph{Phys. Rev. A}, 96:\penalty0 042118, Oct 2017.
\newblock \doi{10.1103/PhysRevA.96.042118}.

\bibitem[Kolkowitz et~al.(2016)Kolkowitz, Pikovski, Langellier, Lukin,
  Walsworth, and Ye]{PhysRevD.94.124043}
S.~Kolkowitz, I.~Pikovski, N.~Langellier, M.~D. Lukin, R.~L. Walsworth, and
  J.~Ye.
\newblock Gravitational wave detection with optical lattice atomic clocks.
\newblock \emph{Phys. Rev. D}, 94:\penalty0 124043, Dec 2016.
\newblock \doi{10.1103/PhysRevD.94.124043}.

\bibitem[Derevianko(2016)]{Derevianko_2016}
Andrei Derevianko.
\newblock Atomic clocks and dark-matter signatures.
\newblock \emph{Journal of Physics: Conference Series}, 723:\penalty0 012043,
  Jun 2016.
\newblock \doi{10.1088/1742-6596/723/1/012043}.

\bibitem[Grotti et~al.(2018)Grotti, Koller, Vogt, H{\"{a}}fner, Sterr, Lisdat,
  Denker, Voigt, Timmen, Rolland, Baynes, Margolis, Zampaolo, Thoumany,
  Pizzocaro, Rauf, Bregolin, Tampellini, Barbieri, Zucco, Costanzo, Clivati,
  Levi, and Calonico]{Grotti2018}
Jacopo Grotti, Silvio Koller, Stefan Vogt, Sebastian H{\"{a}}fner, Uwe Sterr,
  Christian Lisdat, Heiner Denker, Christian Voigt, Ludger Timmen, Antoine
  Rolland, Fred~N. Baynes, Helen~S. Margolis, Michel Zampaolo, Pierre Thoumany,
  Marco Pizzocaro, Benjamin Rauf, Filippo Bregolin, Anna Tampellini, Piero
  Barbieri, Massimo Zucco, Giovanni~A. Costanzo, Cecilia Clivati, Filippo Levi,
  and Davide Calonico.
\newblock {Geodesy and metrology with a transportable optical clock}.
\newblock \emph{Nature Physics}, 14\penalty0 (5):\penalty0 437--441, may 2018.
\newblock ISSN 17452481.
\newblock \doi{10.1038/s41567-017-0042-3}.

\bibitem[McGrew et~al.(2018)McGrew, Zhang, Fasano, Sch{\"{a}}ffer, Beloy,
  Nicolodi, Brown, Hinkley, Milani, Schioppo, Yoon, and Ludlow]{McGrew2018}
W.~F. McGrew, X.~Zhang, R.~J. Fasano, S.~A. Sch{\"{a}}ffer, K.~Beloy,
  D.~Nicolodi, R.~C. Brown, N.~Hinkley, G.~Milani, M.~Schioppo, T.~H. Yoon, and
  A.~D. Ludlow.
\newblock {Atomic clock performance enabling geodesy below the centimetre
  level}.
\newblock \emph{Nature}, 564\penalty0 (7734):\penalty0 87--90, dec 2018.
\newblock ISSN 14764687.
\newblock \doi{10.1038/s41586-018-0738-2}.

\bibitem[Barbiero et~al.(2020)Barbiero, Tarallo, Calonico, Levi, Lamporesi, and
  Ferrari]{Barbiero2020}
Matteo Barbiero, Marco~G. Tarallo, Davide Calonico, Filippo Levi, Giacomo
  Lamporesi, and Gabriele Ferrari.
\newblock {Sideband-Enhanced Cold Atomic Source for Optical Clocks}.
\newblock \emph{Physical Review Applied}, 13\penalty0 (1):\penalty0 014013, jan
  2020.
\newblock ISSN 23317019.
\newblock \doi{10.1103/PhysRevApplied.13.014013}.

\bibitem[Hadley(1986)]{Hadley1986}
G.~Ronald Hadley.
\newblock {Injection Locking of Diode Lasers}.
\newblock \emph{IEEE Journal of Quantum Electronics}, 22\penalty0 (3):\penalty0
  419--426, 1986.
\newblock ISSN 15581713.
\newblock \doi{10.1109/JQE.1986.1072979}.

\bibitem[Shimada et~al.(2013)Shimada, Chida, Ohtsubo, Aoki, Takeuchi, Kuga, and
  Torii]{Shimada2013}
Yosuke Shimada, Yuko Chida, Nozomi Ohtsubo, Takatoshi Aoki, Makoto Takeuchi,
  Takahiro Kuga, and Yoshio Torii.
\newblock {A simplified 461-nm laser system using blue laser diodes and a
  hollow cathode lamp for laser cooling of Sr}.
\newblock \emph{Review of Scientific Instruments}, 84\penalty0 (6):\penalty0
  063101, jun 2013.
\newblock ISSN 00346748.
\newblock \doi{10.1063/1.4808246}.

\bibitem[Hosoya et~al.(2015)Hosoya, Miranda, Inoue, and Kozuma]{Hosoya2015}
Toshiyuki Hosoya, Martin Miranda, Ryotaro Inoue, and Mikio Kozuma.
\newblock {Injection locking of a high power ultraviolet laser diode for laser
  cooling of ytterbium atoms}.
\newblock \emph{Review of Scientific Instruments}, 86\penalty0 (7), jul 2015.
\newblock ISSN 10897623.
\newblock \doi{10.1063/1.4927132}.

\bibitem[Yu et~al.(2006)Yu, Kohel, Kellogg, and Maleki]{Yu2006}
N.~Yu, J.~M. Kohel, J.~R. Kellogg, and L.~Maleki.
\newblock {Development of an atom-interferometer gravity gradiometer for
  gravity measurement from space}.
\newblock \emph{Applied Physics B: Lasers and Optics}, 84\penalty0
  (4):\penalty0 647--652, sep 2006.
\newblock ISSN 09462171.
\newblock \doi{10.1007/s00340-006-2376-x}.

\bibitem[Stanczyk et~al.(2018)Stanczyk, Kafar, Grzanka, Sarzynski, Mroczynski,
  Najda, Suski, and Perlin]{Stanczyk:18}
Szymon Stanczyk, Anna Kafar, Szymon Grzanka, Marcin Sarzynski, Robert
  Mroczynski, Steve Najda, Tadeusz Suski, and Piotr Perlin.
\newblock 450 nm {(Al,In)GaN} optical amplifier with double j-shape waveguide
  for master oscillator power amplifier systems.
\newblock \emph{Opt. Express}, 26\penalty0 (6):\penalty0 7351--7357, Mar 2018.
\newblock \doi{10.1364/OE.26.007351}.

\bibitem[Najda et~al.(2018)Najda, Perlin, Suski, Marona, Stanczyk, Wisniewski,
  Czernecki, Schiavon, and Leszczyński]{10.1117/12.2286044}
S.~P. Najda, P.~Perlin, T.~Suski, L.~Marona, S.~Stanczyk, P.~Wisniewski,
  R.~Czernecki, D.~Schiavon, and M.~Leszczyński.
\newblock {GaN laser diodes for high-power optical integration and quantum
  technologies}.
\newblock In Jen-Inn Chyi, Hiroshi Fujioka, and Hadis Morkoç, editors,
  \emph{Gallium Nitride Materials and Devices XIII}, volume 10532, pages 72 --
  78. International Society for Optics and Photonics, SPIE, 2018.
\newblock \doi{10.1117/12.2286044}.

\bibitem[Saxberg et~al.(2016)Saxberg, Plotkin-Swing, and Gupta]{Saxberg2016}
Brendan Saxberg, Benjamin Plotkin-Swing, and Subhadeep Gupta.
\newblock Active stabilization of a diode laser injection lock.
\newblock \emph{Review of Scientific Instruments}, 87\penalty0 (6):\penalty0
  063109, 2016.
\newblock \doi{10.1063/1.4953589}.

\bibitem[Baillard et~al.(2006)Baillard, Gauguet, Bize, Lemonde, Laurent,
  Clairon, and Rosenbusch]{Baillard2006}
X.~Baillard, A.~Gauguet, S.~Bize, P.~Lemonde, Ph~Laurent, A.~Clairon, and
  P.~Rosenbusch.
\newblock {Interference-filter-stabilized external-cavity diode lasers}.
\newblock \emph{Optics Communications}, 266\penalty0 (2):\penalty0 609--613,
  oct 2006.
\newblock ISSN 00304018.
\newblock \doi{10.1016/j.optcom.2006.05.011}.

\bibitem[Martin et~al.(2016)Martin, Baus, and Birkl]{Martin2016}
Alexander Martin, Patrick Baus, and Gerhard Birkl.
\newblock {External cavity diode laser setup with two interference filters}.
\newblock \emph{Applied Physics B: Lasers and Optics}, 122\penalty0
  (12):\penalty0 1--6, dec 2016.
\newblock ISSN 09462171.
\newblock \doi{10.1007/s00340-016-6575-9}.

\bibitem[Hirano and Ito(2005)]{Hirano2005}
I.~Hirano and N.~Ito.
\newblock {Spectral characteristics of cascade master/slave/slave
  injection-locking of laser diodes}.
\newblock \emph{Optics and Laser Technology}, 37\penalty0 (1):\penalty0 81--86,
  feb 2005.
\newblock ISSN 00303992.
\newblock \doi{10.1016/j.optlastec.2004.03.014}.

\bibitem[Schkolnik et~al.(2020)Schkolnik, Williams, and Yu]{Schkolnik2020}
Vladimir Schkolnik, Jason~R. Williams, and Nan Yu.
\newblock {Direct semiconductor diode laser system for an optical lattice clock
  based on neutral strontium for future tests of fundamental physics in space}.
\newblock In Selim~M. Shahriar and Jacob Scheuer, editors, \emph{Optical,
  Opto-Atomic, and Entanglement-Enhanced Precision Metrology II}, volume 11296,
  page~95. SPIE, feb 2020.
\newblock ISBN 9781510633551.
\newblock \doi{10.1117/12.2549377}.

\end{thebibliography}


\end{document}